\documentclass[journal]{juan}
\usepackage[T1]{fontenc}
\usepackage[latin1]{inputenc}
\usepackage{longtable} 
\usepackage{supertabular}
\usepackage{graphicx}
\usepackage{amssymb}
\usepackage{amsmath}
\usepackage{units}
\usepackage{natbib}
\usepackage{version}
\usepackage{color}
\usepackage{float}
\usepackage{epsfig}
\usepackage{longtable}
\usepackage{supertabular}

\def\cl{C_{\ell}}

\def\be{\begin{equation}}
\def\ee{\end{equation}}
\def\bi{\begin{itemize}}
\def\ei{\end{itemize}}

\begin{document}

\title{A characterization of the diffuse Galactic emissions at large angular scales using the Tenerife data} 

\author{  J.~F.~Mac\'{\i}as-P\'erez$^{1,2}$, R.D.~Davies$^{2}$,  R.~Watson$^{2}$, C.M.~Gutierrez$^{3}$, R.~Rebolo$^{3}$}
\maketitle
\noindent \thanks{(1) LPSC, Universit\'e Joseph Fourier Grenoble 1, CNRS/IN2P3, Institut National Polytechnique de Grenoble, 53 avenue des Martyrs, 38026 Grenoble cedex, France}\\
\thanks{(2) Jodrell Bank Centre for Astrophysics, Alan Turing Building, School of Physics and Astronomy, The University of Manchester, Oxford Road, Manchester, M13 9PL, U.K.\\
\thanks{(2) Instituto de Astrof\'{\i}sica de Canarias, C/ V\'{\i}a L\'actea s/n, La Laguna, 65	Tenerife, Spain \\

\abstract The Anomalous microwave emission (AME) has been proved to be an important component of the Galactic diffuse emission in the range from 20 to 60~GHz.
To discriminate between different models of AME low frequency microwave data from 10 to 20~GHz are needed.  We present here a re-analysis of published and un-published Tenerife
data from 10 to 33~GHz at large angular scales (from 5 to 15 degrees). We cross-correlate the Tenerife data to templates of the main Galactic diffuse emissions: synchrotron, free-free and thermal dust.
We find evidence of dust correlated emission in the Tenerife data that could be explained as spinning dust grain emission. \\ 

\noindent \keywords ISM: general -- ISM: clouds -- Methods: data analysis -- Cosmology: observations -- Submillimeter -- Catalogs 

\date{\today}







\section{Introduction}

\indent The anomalous microwave emission (AME) is an important contributor of the Galactic diffuse emission in the range from
20 to 60~GHz. It was first identified by~\citep{leitch} as
 free-free emission from electrons with temperature, $T_e > 10^6$K. \cite{draine1998}
argued that AME may result from electric dipole radiation due to small
rotating grains, the so-called \emph{spinning dust}. Models of the \emph{spinning dust} emission \citep{draine1998b} 
show that an emissivity spectrum peaking at around 20-50~GHz is able to reproduce the observations~\citep{finkbeiner2003,costa2004, watson, iglesias2005, cassasus, casassus2008,dickinson2009, tibbs}. The initial \emph{spinning dust}  model has been
refined regarding the shape and rotational properties of the dust
grains \citep{ali, hoang2010, hoang2011, silsbee}. An alternative explanations of AME was proposed by
\cite{draine1999} based on magnetic dipole radiation arising from hot
ferromagnetic grains. 
Observations have placed limits of a few per cent on the fractional polarization towards AME targets \citep{batistelli, cassasus, kogut, mason, lopez}. This excludes perfectly-aligned single-domain magnetic grains, however other alignments and grain compositions produce similarly low levels of polarization \citep{2012arXiv1205.7021D}. \\

.

\noindent A correlation between microwave and infrared maps, mainly dominated by dust thermal emission \citep{desert90},
was observed for various experiments, for example
on COBE/DMR~\citep{kogut96a, kogut96b}, OVRO~\citep{leitch},
Saskatoon~\citep{costa1997}, survey at 19~GHz~\citep{costa1998},
Tenerife~\citep{costa1999}. A similar signal was find in compact regions
by~\citep{finkbeiner2003} and in some molecular clouds based on data from
COSMOSOMAS~\citep{genova, watson}, AMI~\citep{scaife1, scaife2},
CBI~\citep{cassasus, castellanos}, VSA~\citep{tibbs} and
Planck~\citep{early}. A recent study of the Small Magellanic Cloud also claims a detection of AME 
\citep{bot}.\\

\indent Independently, \cite{bennett} proposed an alternative explanation of AME based on
flat-spectrum synchrotron emission associated to star-forming
regions to explain part of the WMAP first-year observations. This
hypothesis seems to be in disagreement with results from~\cite{costa2004,fernandez, hildebrandt2007, ysard} which showed that spinning dust 
best explained the excess below 20~GHz. Furthermore,
\cite{davies} showed the existence of important correlation between microwave and
infrared emission in regions outside star-forming areas. More recently,
\cite{kogut2011} discussed the fact that \emph{spinning dust} fits better to
ARCADE data (3.8 and 10~GHz)  than a flat-spectrum synchrotron.\\

To discriminate between the different AME models and from alternative explanations such as those discussed above
low frequency microwave data in the range from 10 to 20~GHz at different angular resolutions are needed.
Indeed in this frequency range we expect the AME spectrum to be significantly distinct from magnetic dust and flat-spectrum synchrotron.
At this respect the Tenerife data set, from 10 to 33~GHz and at large angular scales (from 5 to 15 degrees), is unique.
We present in the following a re-analysis of these data including previously published data~\cite{2000ApJ...529...47G} and un-published
data since January 1998 to December 2000. The paper is structured as follows. Section~\ref{data} presents the Tenerife data and discuss
the re-processing of these data. In Section~\ref{galemissions}
we describe the main Galactic emission mechanisms and the associated templates used in the analysis.
Section~\ref{resolvedpointsources} discusses the point-source contribution to the Tenerife data. In Section~\ref{crosscorrelation} we present the 
cross correlation between the Tenerife data and the Galactic templates are presented. The main results are discussed in
Section~\ref{discussion}. Finally we draw conclusions in Section~\ref{conclusion}.
 
\section{The Tenerife data}
\label{data}

\begin{figure}
\begin{center}
 \includegraphics[width=8cm,height=20cm]{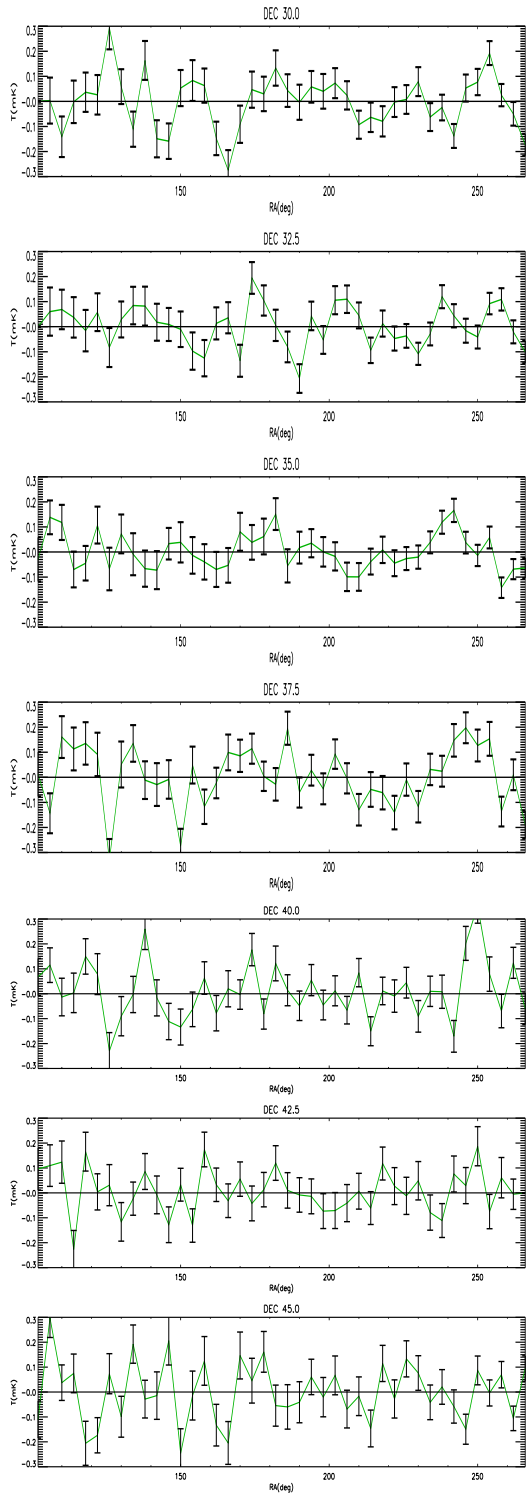} 
\end{center}
\caption[{\bf 10~GHz stacked scans}]
{{\bf 10~GHz stacked scans}. From top to bottom we plot the stacked scans
for the 10~GHz radiometer from declination $30^{\circ}$ to $45^{\circ}$. The data have
been binned into $4^{\circ}$ pixels.}
\label{tenestack10GHz}
\end{figure} 

\begin{figure}
\begin{center}
 \includegraphics[width=8cm,height=20cm]{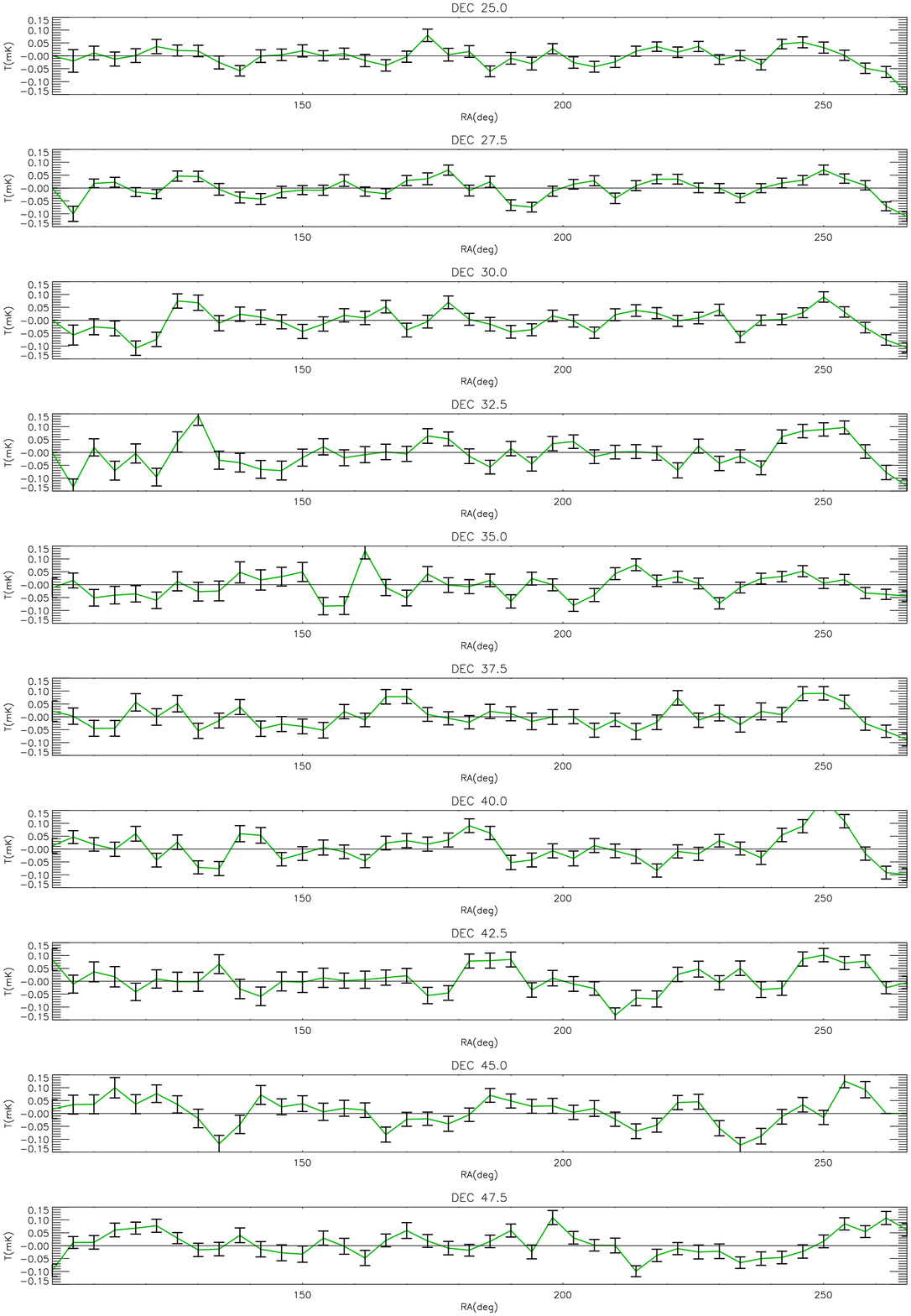} 
\end{center}
\caption[{\bf 15~GHz stacked scans}]
{{\bf 15~GHz stacked scans}. From top to bottom we plot the 15~GHz stacked
scans in the declination range  $25^{\circ}$ to $47^{\circ}.5$. The data have been binned
into $4^{\circ}$ pixels.}
\label{stackedscans15}
\end{figure} 

\begin{figure}
\begin{center}
 \includegraphics[width=8cm,height=6cm]{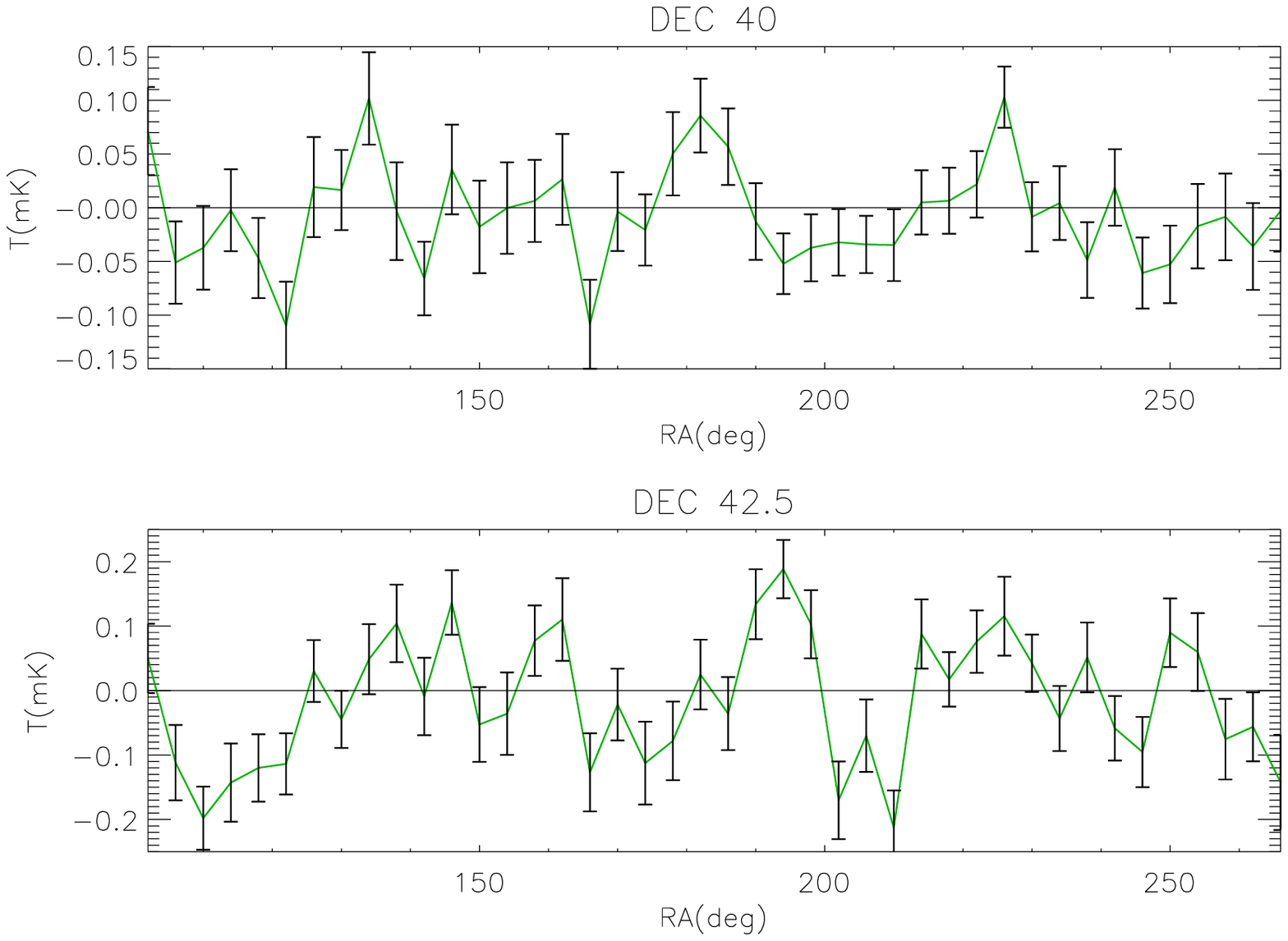} 
\end{center}
\caption[{\bf 33~GHz stacked scans}]
{{\bf 33~GHz stacked scans}. From top to bottom we plot the stacked scans
for the 33~GHz radiometer in the triple beam configuration for declinations 
$40^{\circ}$ to $42.5^{\circ}$. The data have been smoothed into $4^{\circ}$ pixels.}
\label{stackedscans33}
\end{figure}

The Tenerife experiment observed unidimensional scans at constant declination at 10, 15 and
33~GHz using a triple beam pattern of FWHM $5^{\circ}$ and of $8.1^{\circ}$ beam spacing. The region of the sky centered at declination $40^{\circ}$ was chosen for observations
because it corresponds to the largest area of the sky at high latitudes 
where contamination from foregrounds is a minimum.
To reconstruct 2D maps of the sky, consecutive declinations separated by half the beamwidth
($2.5^{\circ}$) are observed. Each single declination is repeatedly observed
until sufficient sensitivity and full RA coverage is achieved. 
The scheduling of the observations
takes into account the position of the Sun so that its contribution to the data
is a minimum. This requires observations of the same declination at different times of the year 
for full RA coverage. Observations were performed day and night. Day-time observations
present an increase in total power and noise with respect to night-time observations due
to receiver gain changes.  In extreme cases, data observed at day time have to be removed. \\

The atmospheric contribution to the data depends on frequency, being severe at 33~GHz
and relatively small at 10~GHz. The observing efficiency depends mainly on the weather
conditions. Data strongly affected by atmospheric contamination cannot be used.
In addition, technical problems such as power cutoff, warmed and oscillating HEMT amplifiers,
RF interference, failure in the electronic systems, etc. can also affect the data.
At 10 and 15~GHz more than $80\%$ of the observed data are useful. However, at 33~GHz only
about $10\%$ of data are kept, due mainly to atmospheric effects.  \\

We use in the following the full Tenerife data set which includes new data
(taken from January 1998 to December 200) with respect to previously published releases
(see for example \cite{2000ApJ...529...47G,1994Natur.367..333H} ).
These data have been completely reprocessed including hand editing to remove clearly systematic contaminated
regions and accurate removal of the atmospheric emission using a MEM based baseline removal technique as
described in~\cite{2000ApJ...529...47G}.  The latter has been improved to enlarge the sky region for which atmospheric
residuals are negligible. In particular, this allows us to consider in the following analysis regions at
low Galactic latitudes. Figures~\ref{tenestack10GHz}, \ref{stackedscans15} and \ref{stackedscans33} present the
cleaned Tenerife data at 10, 15 and 33~GHz respectively. The main properties of these clean
data: central declination of the scan, mean temperature, mean noise per beam and r.m.s per 1$^{\circ}$
pixel in the RA interval $150^{\circ}$ to $250^{\circ}$ are presented in Tables~\ref{10GHzdatastats},\ref{15GHzdatastats} and \ref{33GHzdatastats}
for the 10, 15 and 33~GHz channels respectively.

\begin{table}
\caption[10~GHz data statistics.]
{{\bf 10~GHz data}. Mean temperature, mean noise per beam and r.m.s per 1$^{\circ}$
pixel for the 10~GHz final stacks in the RA interval $150^{\circ}$ to $250^{\circ}$.}  
\vspace{0.3cm}
\begin{tabular}{c c c c}
Declination (deg) & Mean T ($\mu$ K)  & $\sigma$($\mu$ K) & r.m.s ($\mu$K) \\
\hline
\hline
      30.0 &    -3.7  &     52.9 &     150.5  \\
      32.5 &    -5.0  &    45.9  &    147.8  \\
      35.0 &     10.0 &     47.1 &     140.5  \\
      37.5 &     7.5  &    51.0  &    153.1   \\
      40.0 &     13.6 &     49.6 &     147.4  \\
      42.5 &     7.9  &    54.1  &    146.0  \\
      45.0 &    -7.8  &    59.2  &    215.5  \\
\end{tabular}
\label{10GHzdatastats}
\end{table}

\begin{table}
\caption[15~GHz data statistics]
{{\bf 15~GHz data}. Mean, mean noise per beam and r.m.s per 1$^{\circ}$ pixel for the 15~GHz
final stacks in the RA interval $150^{\circ}$ to $250^{\circ}$.}  
\vspace{0.3cm}
\begin{tabular}{c c c c}
Declination & Mean T ($\mu$ K) & $\sigma$($\mu$ K) & r.m.s ($\mu$K) \\
\hline
\hline
      25.0 &    3.6  &   18.1 &    51.7 \\
      27.5 &    4.6  &   16.3 &    48.7 \\
      30.0 &    3.4  &   19.3 &    61.1 \\
      32.5 &    3.9  &   22.7 &    67.2 \\
      35.0 &   0.25  &   20.6 &    74.6 \\
      37.5 &    8.3  &   22.4 &    78.6 \\
      40.0 &    7.8  &   20.7 &    75.7 \\
      42.5 &    5.0  &   22.6 &    81.3 \\
      45.0 &   -8.2  &   21.3 &    75.8 \\
      47.5 &   -5.9  &   19.3 &    70.5 \\
\end{tabular}
\label{15GHzdatastats}
\end{table}

\begin{table}
\caption[33~GHz data statistics.]
{{\bf 33~GHz data}. Mean, mean noise per beam and r.m.s per 1$^{\circ}$ pixel for the 33~GHz
final stacks in the RA interval $150^{\circ}$ to $250^{\circ}$.}  
\vspace{0.3cm}
\begin{tabular}{c c c c}
Declination & Mean T ($\mu$K)  & $\sigma$($\mu$K) & r.m.s ($\mu$K) \\
\hline
\hline
      40.0 &     -3.1 &      27.2 &      80.8 \\
      42.5 &      0.2 &      43.7 &      167.7  \\
\hline
\end{tabular}
\label{33GHzdatastats}
\end{table}

\begin{figure*}
\begin{center}
 \includegraphics[angle=90,width=16cm,height=7cm]{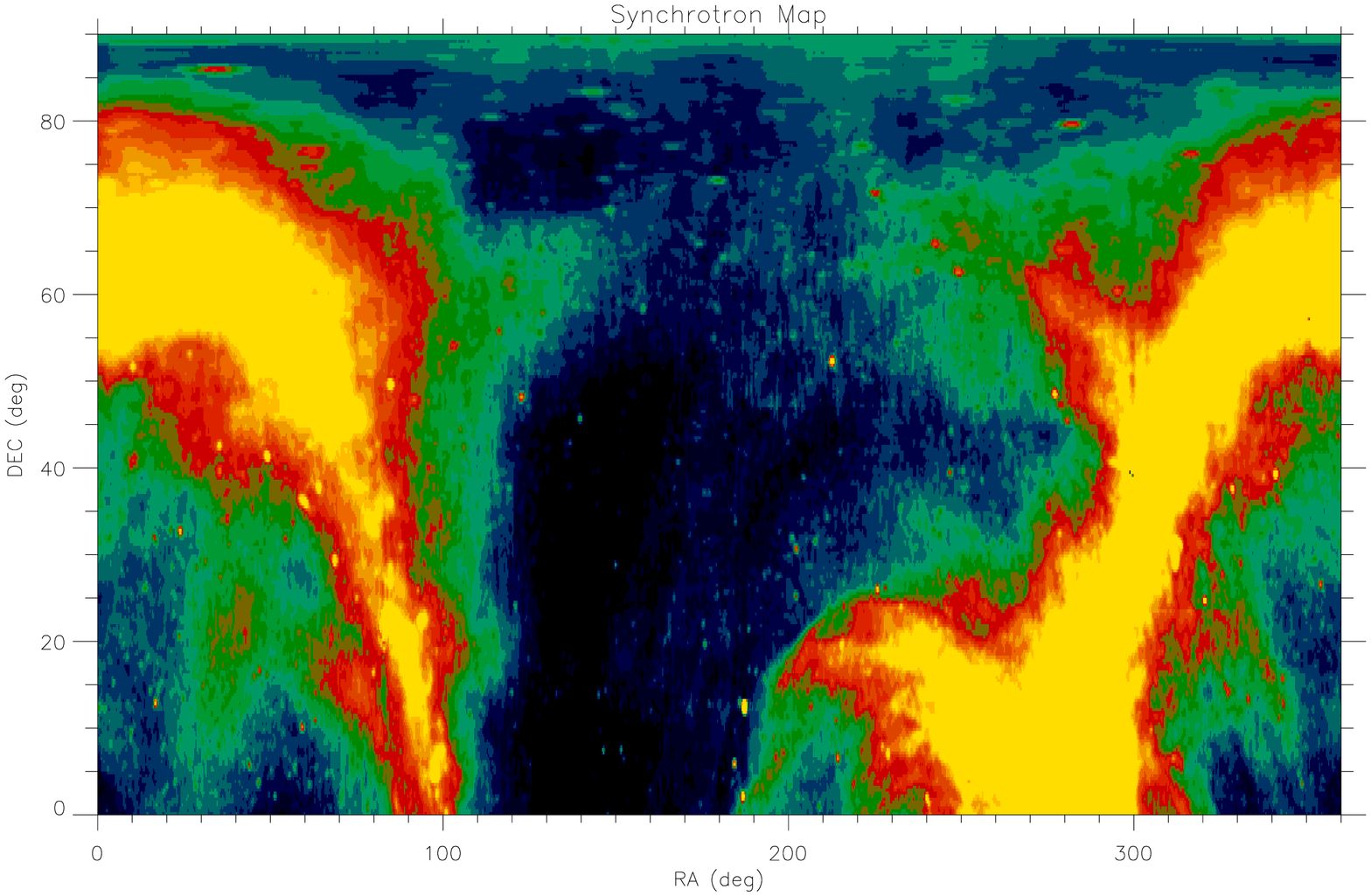}
 \includegraphics[angle=90,width=16cm,height=7cm]{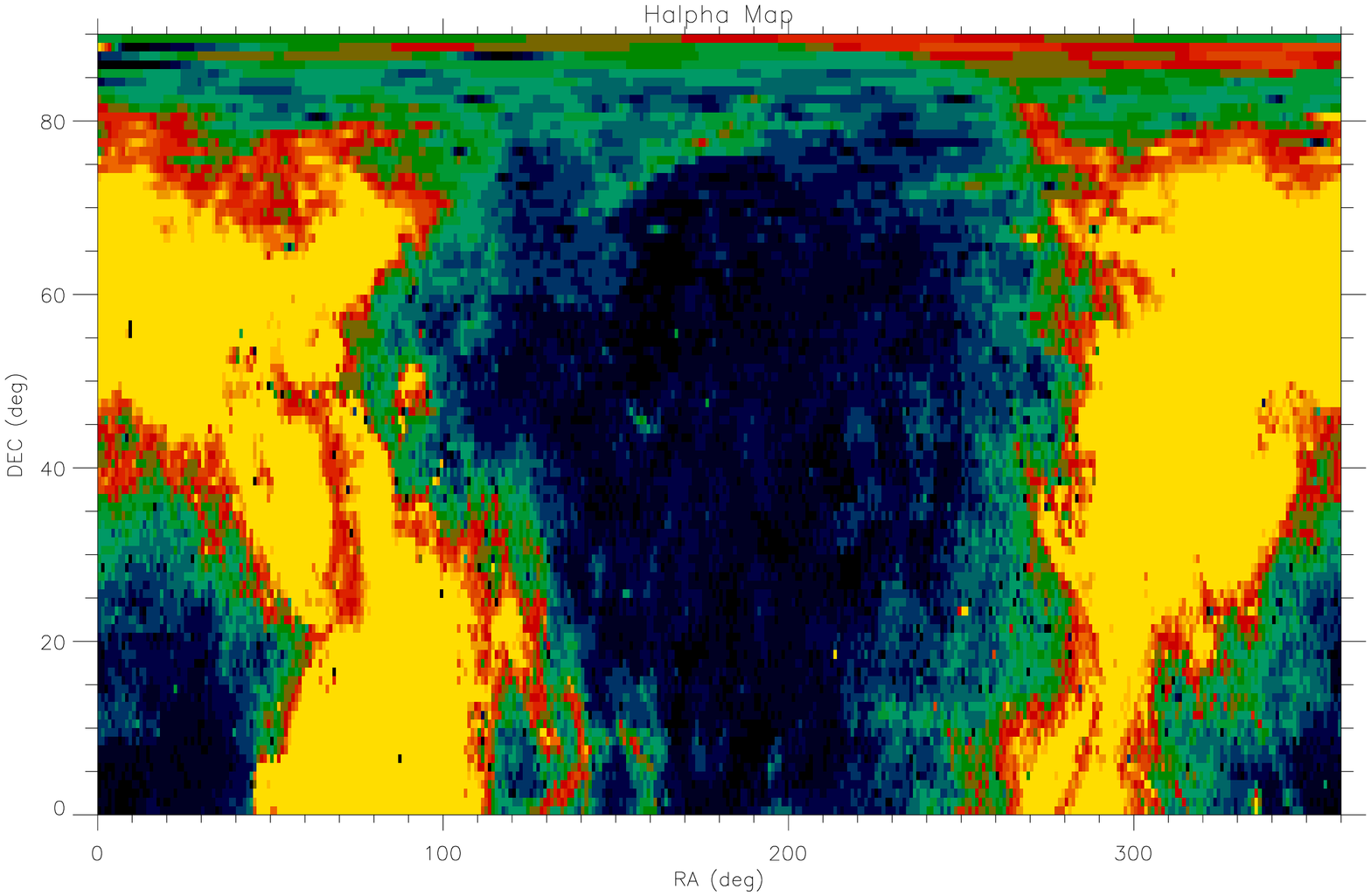}
 \includegraphics[angle=90,width=16cm,height=7cm]{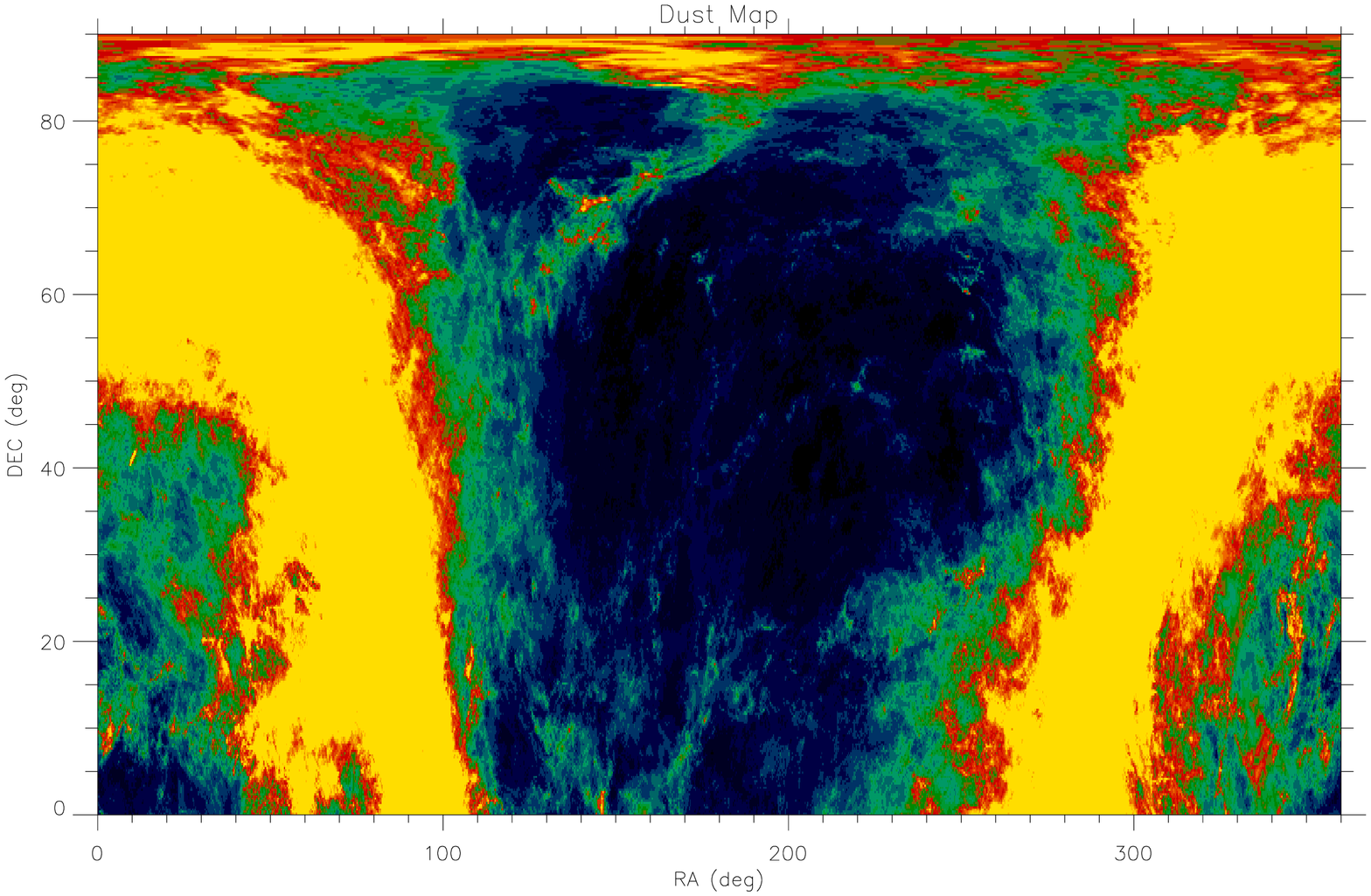}
\end{center}
\caption[{\bf Maps of the main Galactic diffuse emissions.}]
{From top to bottom. {{\bf $408$ MHz synchrotron map.} Resolution: $0^{\circ}.85$ Pixels: $0^{\circ}.33$.
{\bf WHAM H$\alpha$ map.} This map is courtesy of the WHAM collaboration. Resolution: $1^{\circ}$ Pixels: $1^{\circ}$.
\bf Combined DIRBE+IRAS $100 \mu$m dust map.} Resolution: $6$ arcmin Pixels: $0^{\circ}.25$.   }
\label{galemissionmaps}
\end{figure*}

\section{Galactic diffuse emission.}
\label{galemissions}
We present in this section the main known Galactic diffuse mechanisms: synchrotron, free-free,
and vibrational and rotational dust.

\subsection{Synchrotron}
\index{s@{\bf S}!synchrotron}
Synchrotron emission results from cosmic-ray electrons accelerated in magnetic fields, and
thus, depends on the energy spectrum of the electrons and the intensity of the magnetic field
\citep{ribicki,longair}. The local energy spectrum of the electrons has been
measured to be, for energies contributing to the observed radio synchrotron emission, a 
power law to good approximation with index from about 
-2.7 to -3.3 over this energy range \citep{1999tkc..conf..204D}. Such an increase of energy 
spectrum slope is expected, as the energy loss mechanism for electrons increases
as the square of the energy. 

Radio surveys at frequencies less than 2~GHz are dominated by synchrotron emission
\citep{1987MNRAS...225...307}. The only all-sky survey available at these frequencies
is the 408 MHz map \citep{1981A&A...100..209H}. This survey has a resolution of $0^{\circ}.85$
and was produced using the Parkes 64-m telescope in Australia for the southern sky and the
Bonn 100m and Jodrell Bank MK1A telescopes for the northern sky. The scanning strategy with the
Bonn telescope was to fix the azimuth at the local meridian and scan up and down in
elevation at a rate of about $6^{\circ}/$min. This technique reduced
the atmospheric contribution to the map but led to a set of vertical stripes (constant RA)
separated by $7^{\circ}$. The quoted errors in the temperature scale are of the order of $10 \%$
and $\pm 3$ K in the absolute brightness temperature levels. \\

In addition there is the 1420 MHz survey \citep{1988A&AS...74....7R}
which covers the declination range $-19^{\circ}$ to $90^{\circ}$ and has a FWHM of 
$0^{\circ}.58$. Stripes are also present in this map due to the scanning strategy which consisted of 
azimuthal scans at constant elevation.
The errors in the temperature scale are of the order of $5 \%$ and $\pm 0.5$ K in the absolute 
brightness temperature.

The 408 MHz and 1420 MHz maps have been used to determine the synchrotron spectral index
at radio frequencies \citep{1988A&AS...74....7R,1996MNRAS.278..925D}. Assuming 
$T_{\nu} \propto \nu^{-\beta}$ and after destriping and correction for zero levels,
spectral indexes, $\beta$, in the range $2.8$ to $3.2$ were found. The spatial angular power spectrum
of the synchrotron emission $\cl^{sync}$ is poorly understood but is believed to be 
$\cl^{sync} \propto \ell^{-3}$ \citep{1999NewA....4..443B}. At high Galactic latitudes in
the region observed by the Tenerife experiment the synchrotron spatial angular power spectrum
is slightly flatter than $l^{-2}$ \citep{lasenbymoriond}. 
In the following sections, we will use the destriped version of the 408 MHz map 
\citep{1996MNRAS.278..925D} as  a template of the synchrotron emission.   This map is shown
on Figure~\ref{galemissionmaps}.

\subsection{Free-Free}
\index{f@{\bf F}!free free}
When a charged particle is accelerated in a Coulomb field it will emit radiation which is called
braking radiation or Bremsstrahlung.
The Galactic free-free emission is the thermal bremsstrahlung from hot electrons ( $\sim 10^{4}$K )
produced in the interstellar gas by the UV radiation field \citep{1999tkc..conf..204D}. 
This emission is not easily identified at radio frequencies, except near the Galactic plane.
At higher latitudes it must be separated from synchrotron emission 
by virtue of their different spectral indices, since the spectral index of optically thin free-free emission is $\beta^{ff} = 2.1$. \\

The diffuse Galactic recombination radiation, H$_{\alpha}$ is a good tracer of free-free emission 
since both are emitted in the same Warm Ionised Medium (WIM) and both have 
intensities proportional to the {\it emission measure},
\begin{equation}
EM = \int n_{e} n_{p} d\ell \simeq \int n_{e}^{2} d\ell
\end{equation}  
The ratio of free-free brightness temperature $T_{b}^{ff}$ to the H$_{\alpha}$ surface brightness
$I_{\alpha}$ in R (Rayleigh) is \citep{1998PASA...15..111V}
\begin{equation}
\frac{T_{b}^{ff} [mK]}{I_{\alpha}[R]} = 10.4 \nu^{-2.14} T_{4}^{0.527} 10^{0.029/T_{4}} (1+0.08)
\label{halphatotempratio}
\end{equation}
where $\nu$ is the observing frequency in GHz, $T_{4}$ is the temperature of the electrons in units
of 10$^{4}$ K
, and the last factor $0.08$ corresponds to helium which is assumed completely ionised and creates
free-free emission like hydrogen but does not emit H$_{\alpha}$ light. 

Recently a full sky survey of  H$_{\alpha}$ light has been released by the WHAM 
(Wisconsin H$_{\alpha}$ Mapper) collaboration \citep{1999ApJ...523..223H}. 
The WHAM instrument consists of a 6 inch dual-etalon Fabry-Perot
spectrometer with a narrow band filter of FWHM $\sim 20$ \AA which images onto a
cryogenically cooled 1024 x 1024 CCD. The resolution of the survey is $\sim 1^{\circ}$. 
The spatial power spectrum of free-free emission $\cl^{ff}$ has not yet been determined
from the WHAM map however analysis of H$_{\alpha}$ images of the North Celestial
Pole Area \citep{1997ppeu} suggests  $\cl^{ff} \propto \ell^{-2.27 \pm 0.07}$.

In Figure~\ref{galemissionmaps} we present a map of the northern sky produced by the WHAM 
survey. In the following sections we will use this map as a template for the free-free emission.
At intermediate Galactic latitudes (say $|b| > 30^{\circ}$) about 10 $\%$ of 
the H$_{\alpha}$ light is absorbed by dust and
therefore estimates of free-free emission from H$_{\alpha}$ will be systematically lower
than the true value. At latitudes below 10$^{\circ}$ this correction becomes increasingly
uncertain.

\subsection{Vibrational Dust}
\index{d@{\bf D}!dust!vibrational}
At the higher frequency range ( $\geq$ 100~GHz) of the microwave background experiments, 
dust emission starts to
become dominant. Dust grains are heated by interstellar radiation, absorbing optical and
UV photons and emitting energy in the far infrared.

The intensity of the emission from an ensemble of dust grains is given by
\begin{equation}
I(\nu) = \int \epsilon(\nu) d\ell
\end{equation}
where $\epsilon(\nu)$ is the emissivity at frequency $\nu$, and the integral is along
the line of sight. In the Rayleigh--Jeans regime and assuming a constant line-of-sight
density of dust,
\begin{equation}
T_{b} \propto \epsilon(\nu) \nu^{-2}
\end{equation}
where $T_{b}$ is the brightness temperature of the dust emission.

The spectrum of the dust emission has
been measured at millimeter and submillimeter wavelengths by the Far-Infrared Absolute Spectrophotometer
(FIRAS) and can be fitted by a single greybody spectrum of temperature $17.5$ K 
and emissivity $\propto \nu^{2}$ \citep{1996A&A...312..256B} at high Galactic latitudes. From 
IRAS observations of dust emission \citep{1992BAAS...24.1120G}, it was found that
the spatial power spectrum of the dust fluctuations is $\cl^{dust} \propto \ell^{-3}$. This has
also been confirmed at larger angular scales by the COBE-DIRBE satellite. 

In the following sections we will use the combined IRAS-DIRBE map at 100 $\mu$m 
\citep{1998ApJ...500..525S} as a template for the dust emission. This map has resolution of
FWHM $\sim 6$ arcmin and covers the full sky. Zodiacal light and point sources have been removed 
from the map and the regions of the sky which were not observed by the IRAS satellite have
been replaced by DIRBE data. The combined map preserves the DIRBE zero point and calibration.
This map is in units of MJy/sr. In Figure~\ref{galemissionmaps} we present the northern sky part of the combined IRAS-DIRBE map.

\subsection{Rotational Dust.}
\index{d@{\bf D}!dust!rotational}
Small spinning interstellar dust grains containing $10^{2}$--$10^{3}$ atoms can produce detectable 
rotational emission in the 10--100~GHz range. This emission
depends on the component of the electric dipole moment perpendicular to the angular velocity of 
the grain and on the physical properties of the interstellar medium \citep{1998ApJ...494L..19D}.
For these small grains, rotational excitation is dominated by direct collisions with ions
and {\it plasma drag}. The very smallest grains ($ N \leq 150$) have their rotation damped
primarily by electric dipole emission; for 150 $\leq$ N $\leq 10^{3}$ plasma drag dominates.

In the following sections, we show models for the spectrum of the spinning dust 
emission provided by Dr. Draine  (private communication). These models depend on a large set of parameters such as the distribution
of grain sizes, the charge of the grains, the composition of the grains and the physical
properties of the interstellar medium which were fixed by the authors.
However, the normalization of the model can
be assumed a free parameter although it is related to the hydrogen column density
in the interstellar medium which is of the order of a few times $10^{20}$. 

The spatial distribution of small dust grains is not well-known although it seems 
reasonable to believe that it is not different from that of larger grains but for dense regions
where dust coagulation may deplete small grains. For the
purpose of this work we will use the IRAS-DIRBE 100 $\mu$m map as a template for the
spinning dust grains. 

\section{Galactic and Extra-Galactic point sources.}
\label{resolvedpointsources}
\index{p@{\bf P}!point sources!resolved}
The contribution from resolved point sources to the Tenerife data at 10 and 15~GHz
was extrapolated from data
of the Michigan monitoring program (M. Aller and H. Aller 2000, private communication).
The Michigan program regularly monitors point sources with fluxes above $0.5$ Jy at 4.8, 8.0 and
14.5~GHz. The Michigan catalog is neither complete in flux or time domain. Well-known strong variable
sources such as 3C345 are regularly observed and data at all observing frequencies are available.
Weak sources are poorly observed and often data are available at a single frequency of the three
possible. Further, data are available up to June 1999 while the Tenerife experiment operated
until September 2000. For the 33~GHz Tenerife data we used the Metsahovi catalog (Metsahovi
group, private communication) which regularly monitors sources above $1$ Jy at 22 and
37~GHz. This catalog is neither complete in flux or time although observations are available
up to January 2000.

We have developed software to produce time and frequency complete point source catalogs. The flux of
the sources for each frequency was interpolated in time by fitting Fourier Series 
to the data. If the number of independent observations per frequency was smaller than 10, the
sources were assumed not variable and the data was fitted to a constant with time. 
The extrapolation of the source fluxes into the future time was performed assuming no variability 
since last observed data point. Spectral indices were calculated 
for each source from the available data. If only observations at a single frequency were 
available we assumed a flat spectrum. We constructed four point source catalogs at 10,
13, 15 and 33~GHz covering the time range January 1984 to December 2000 with one Julian
day sampling and a flux limit of about 0.5 Jy.
    
For each single day of observations with the Tenerife radiometers we estimated the total
contribution from point sources to the data. 
This contribution was calculated from the extrapolated catalog
at 10, 15 and 33~GHz and the flux, $S$, was converted into antenna temperature, $T$, using
\citep{rohlfs}
\begin{equation}
T = \frac{S \lambda^{2}}{2 k \Omega}
\end{equation}
where $\Omega$ is the beam solid angle and $\lambda$ the wavelength. The daily point source
contributions were stacked in the same manner as the Tenerife data and subtracted from the
Tenerife stacks. In the left column of figure \ref{15pointsourcecont} we plot the Tenerife stacked
data at 15~GHz for the 10 declinations presented in this paper and overplot in green the contribution 
from point sources to the data. In the right column, we plot the stacks after subtraction of
the point sources. 
\begin{figure*}
\begin{center}
 \includegraphics[width=18cm,height=20cm]{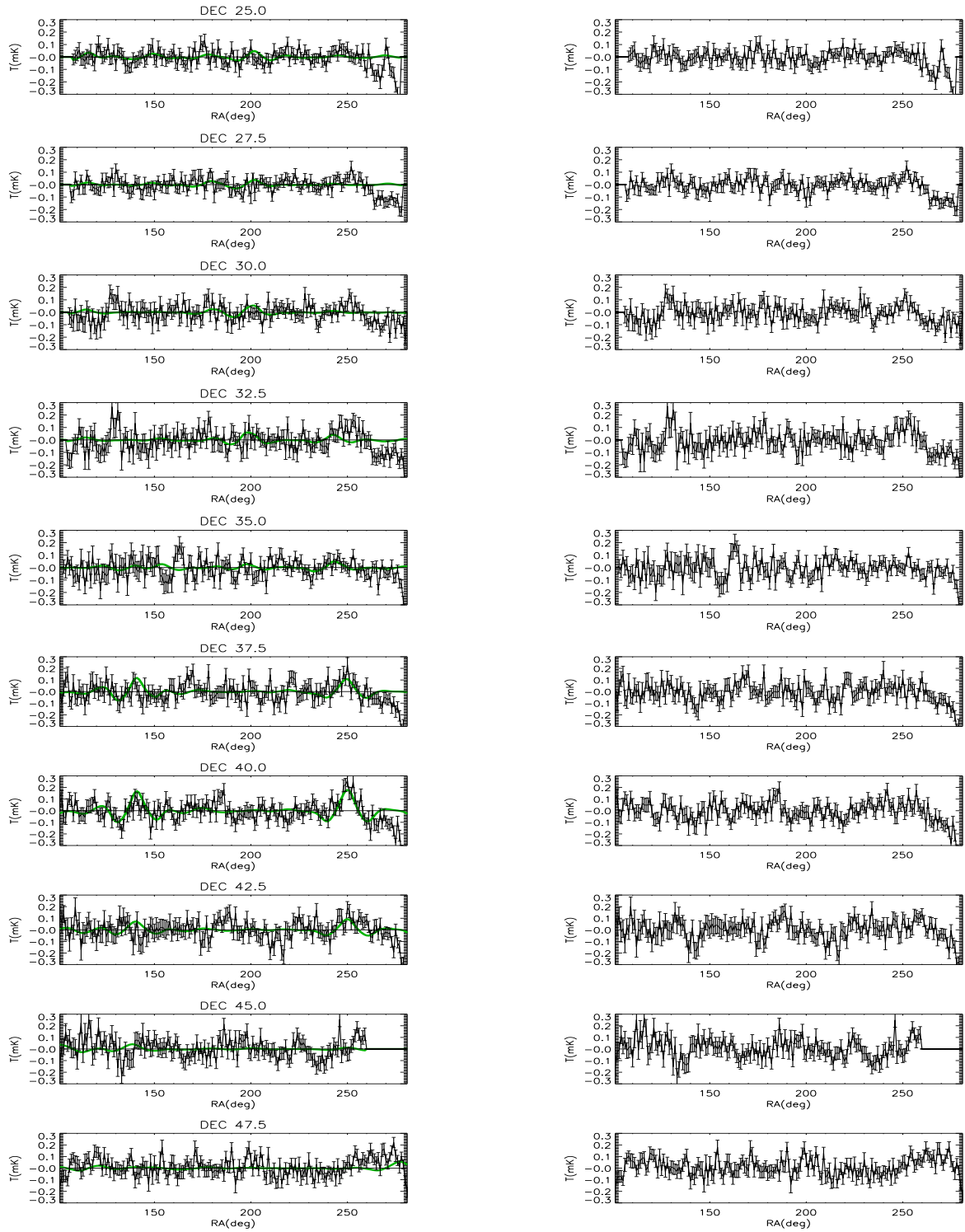}
\end{center}
\caption[{\bf Extrapolated point source contribution to the 15~GHz Tenerife data}]
{{\bf Extrapolated point source contribution to the 15~GHz Tenerife data.} Left column: observed scans.
Right column: point source-corrected scans.}
\label{15pointsourcecont}
\end{figure*}
\subsubsection{Unresolved sources.}
\index{p@{\bf P}!point sources!unresolved} 
A study of the contribution of unresolved point sources (ie. weak point sources not detected
individually) to CMB experiments has been produced
by \cite{1989ApJ...344...35}. They used numerous surveys, including VLA and IRAS data, to put limits on
the contribution to single beam CMB experiments by a random distribution of point sources.
The contribution from unresolved point sources to the Tenerife data deduced from the previous analysis
is presented in table \ref{unresolvedpointsourcestable}. We consider sources with flux under
1 Jy at 5~GHz and a main beam of 5$^{\circ}$.
\begin{table}
\caption{Contribution from unresolved point sources to the Tenerife experiment.}
\vspace{0.5 cm}
\begin{center}
\begin{tabular}{c c}
\hline
$\nu$ (GHz) & $\frac{\Delta T}{T}$ ($\mu$K) \\
\hline
\hline
10  &  $\leq 7.5  \mu$K \\
15  &  $\leq 4.0   \mu$K \\
33  &  $\leq 0.6   \mu$K \\
\hline
\end{tabular}
\end{center}
\label{unresolvedpointsourcestable}
\end{table}

\section{Assessment of the Galactic contribution to the Tenerife data.}
\label{crosscorrelation}
To assess Galactic contribution to the Tenerife data, we correlated
observations with the Galactic synchrotron, dust and H$_{\alpha}$ emission templates described in
the previous sections. We convolved the foreground maps with the Tenerife beam at
each of the Tenerife frequencies before the correlation was performed.
The Tenerife data used in the correlation are the final stacked data presented in 
Section~\ref{data}. At 15~GHz we used ten declination stacks in the range 
$25^{\circ}-47^{\circ}.5$; seven declination stacks at 10~GHz in the range  
$30^{\circ}-45^{\circ}$; and only two declination stacks at 33~GHz covering declinations
$40^{\circ}$ and $42^{\circ}.5$. The data were processed so that data at $|b| \leq 15$ were
excluded from the baseline fit and therefore were neither stacked or reconstructed. 
The discrete point sources were subtracted from the Tenerife stacks as discussed in the previous
section.
\subsection{The Method}
To simultaneously correlate the Tenerife data to the three Galactic templates we use a method
which was first applied to this problem by \cite{1996ApJ...464L..11G} to fit Galactic and
extra-Galactic templates in Fourier space to the COBE-DMR data. This method was
applied to the Tenerife data first by \cite{1999ApJ...527L...9D} and then by 
\cite{2001MNRAS.320..224M} to study the possible emission of spinning dust at the Tenerife
frequencies.

Assuming that the microwave data is a superposition of CMB, noise and Galactic components,
we can write
\begin{equation}
y = a X + x_{CMB} + n
\end{equation}
where y is a Tenerife data vector of N pixels; X is an $N \times M$ element matrix containing 
$M=3$ foreground templates convolved with the Tenerife beam; $a$ is a vector of size $M$ that 
represents the levels at which these foreground templates are present in the Tenerife data -
correlation coefficients for each foreground template; $n$ is the instrumental noise in the
data; and $x_{CMB}$ is the CMB component of the data. For this analysis we assume the
noise and CMB to be uncorrelated.

The minimum variance estimate of $a$ is given by
\begin{equation}
\hat{a}= {\left[X^{T}C^{-1}X\right]}^{-1}X^{T}C^{-1} y
\end{equation}
with errors given by $\sigma_{\hat{a}_{i}}= \sum_{ii}^{1/2}$ where $\sum$ is given as
\begin{equation}
\sum = \left<\hat{a}^{2}\right> - \left<\hat{a}\right>^{2} = {\left[X^{T}C^{-1}X\right]}^{-1}
\end{equation}
In the above, C is the total covariance matrix, the sum of the noise covariance
matrix and CMB covariance matrix. The noise covariance matrix of the Tenerife data is
taken to be diagonal - no correlation between different pixels. The CMB covariance
matrix was obtained analytically following \cite{1998PhDT.........Z}.

This correlation method produces minimum variance and unbiased estimates
of $a$ if the following holds
\begin{itemize}
\item{{\it The error in the Tenerife data is Gaussian and with zero mean}. To a very good 
approximation the instrumental noise in the Tenerife data is uncorrelated.}

\item{{\it The templates perfectly trace the foreground emissions at microwave frequencies}.
There may be components of emission present in the Tenerife data, apart from those we have
identified, for which there are not obvious counterparts at other frequencies. Moreover, 
the templates we use do not perfectly trace
the microwave emission of the Galactic foreground. The latter is specially important
in the case of vibrational and rotational dust for which the template comes from data at much higher
frequencies. For example, \cite{1998ApJ...494L..19D} proposed a $12 \mu$m map would be a much
better tracer of rotational dust emission because small grains will also emit at this frequency.
Also, the H$_{\alpha}$ emission is absorbed at low Galactic
latitudes by interstellar dust and therefore does not fully trace the free-free emission at those
latitudes. Moreover, the 408 MHz map has residual striping.}
\item{{\it The templates are perfect -- error bars equal to zero}. The error estimated for the
templates are about 5 -- 10 \% of the signal. They are not corrected for in this analysis
because they are small compared to the errors in the Tenerife data.}
\item{{\it The templates are not correlated}. It there is correlation between the different
templates the set of minimum variance correlation coefficients  $\hat{a}$ is degenerate and 
therefore we would not be able to discriminate among the different foreground components. At low
Galactic latitudes, the Galactic plane emission dominates and \cite{2001MNRAS.320..224M} have
found the foreground templates are correlated in this region.}
\item{{\it The correlation coefficient $a$ is the same throughout the area of sky for which
the correlation is performed}. If it is not, the error associated with $a$ is systematically 
underestimated. As a double check, we divided the sky observed, into independent
areas and performed the same correlation test in each, calculating a mean correlation
coefficient and the dispersion of the individual values which in most cases was in good
agreement with the error of $a$ calculated for the total area.} 
\end{itemize}
\subsection{Cross-Correlation results.}
The cross-correlation results are presented in table \ref{forecorrresults}.
The correlation was performed for three different Galactic
cuts $b > 20$, $b >30$ and $b >40 $. $\sigma_{gal}$ represents the r.m.s. of the Galactic maps
after convolution with the Tenerife beam.  $\hat{a}$ and $\sigma_{\hat{a}}$ are the correlation
coefficient and the error associated with it. They have units of $\mu$K/K, $\mu$K/(MJy sr$^{-1}$)
and  $\mu$K/R for the correlation with the 408 MHz, 100 $\mu$m and WHAM maps respectively. 
$\Delta T$ is the r.m.s. contribution from
the Galactic foregrounds to the Tenerife data, which was obtained as $\sigma_{gal} \times \hat{a}$.
This analysis is improved with respect to previous analyses by \cite{1999ApJ...527L...9D}
and \cite{2001MNRAS.320..224M} first, because we present Tenerife data at 10 and 15~GHz
for a much larger area of the sky and with improved sensitivity, and second, because we also include 
in the cross-correlation a template for the free-free emission. Moreover, data at 33~GHz have 
also been included in the analysis although the area of the sky observed is significantly 
smaller than at 10 and 15~GHz and consequently the error bars in the estimated cross-correlation 
coefficients much larger.
\begin{table}
\caption[Cross-correlation results]
{{\bf Cross-correlation results}. See text for details}
\vspace{0.5 cm}
{\tiny
\begin{tabular}{|c|c|c|c|c|c|}
\hline
Galactic cut & $\nu$ (GHz) & \multicolumn{2}{c|}{$\sigma_{gal}$} & $\hat{a} \pm \sigma_{\hat{a}}$ & $\Delta T$ ($\mu$K) \\ 

\hline
\hline
                  &        & 408 MHz    & 1.0217  & $17 \pm 6$   & $17.4  \pm 6.0$ \\
\cline{3-6}
                  & 10 GHz & 100 $\mu$m & 0.3256  & $23 \pm 17$  & $7.5 \pm 5.5$ \\
\cline{3-6}
                  &        & WHAM      & 0.2513  & $42 \pm 20$  & $10.5 \pm 5.5$ \\
\cline{2-6}  
                  &        & 408 MHz   & 1.0244  & $9 \pm 3 $   & $9.2 \pm 3.0$ \\
\cline{3-6}
$ b > 20^{\circ}$ & 15 GHz & 100 $\mu$m & 0.3308  & $66 \pm 8$   & $21.8 \pm 2.6$ \\
\cline{3-6}
                  &        & WHAM      & 0.2758  & $-10 \pm 10$ & $-2.7 \pm 2.7$ \\
\cline{2-6}
                  &        & 408 MHz   & 0.8904  & $4 \pm 9 $   & $3.6 \pm 8.0$  \\
\cline{3-6}
                  & 33 GHz & 100 $\mu$m & 0.3089  & $86 \pm 27$  & $26.0 \pm 8.0$  \\
\cline{3-6}
                  &        & WHAM       & 0.2768  & $21 \pm 51$  & $5.8 \pm 10.0$ \\   
\hline 
\hline 
                  &        & 408 MHz    & 0.8722  & $15 \pm 7$   & $13.0 \pm 6.0$  \\
\cline{3-6}
                  & 10 GHz & 100 $\mu$m & 0.2028  & $72 \pm 30$  & $14.6 \pm 6.0$ \\
\cline{3-6}
                  &        & WHAM       & 0.1435  & $83 \pm 42$  & $11.9 \pm 6.0$ \\
\cline{2-6}  
                  &        & 408 MHz    & 0.8544  & $13 \pm 4$   & $11.1 \pm 3.4$ \\
\cline{3-6}
$ b > 30^{\circ}$ & 15 GHz & 100 $\mu$m & 0.2036  & $-1 \pm 14$  & $-0.2 \pm 2.8$ \\
\cline{3-6}
                  &        & WHAM       & 0.1504  & $30 \pm 20$  & $4.5 \pm 3.0$ \\
\cline{2-6}
                  &        & 408 MHz    & 0.7749  & $7 \pm 10$   & $5.4 \pm 7.7$  \\
\cline{3-6}
                  & 33 GHz & 100 $\mu$m & 0.2143  & $90 \pm 45$  & $19.2 \pm 9.6$  \\
\cline{3-6}
                  &        & WHAM       & 0.1473  & $17 \pm 63$  & $2.5 \pm 9.3$ \\   
\hline
\hline
                  &        & 408 MHz    & 0.7676  & $11 \pm 9.5$ & $8.4 \pm 8.0$ \\
\cline{3-6}
                  & 10 GHz & 100 $\mu$m & 0.1898  & $110 \pm 45$ & $20.9 \pm 8.5$ \\
\cline{3-6}
                  &        & WHAM       & 0.1292  & $77 \pm 52$  & $9.9 \pm 6.7$ \\
\cline{2-6}  
                  &        & 408 MHz    & 0.7412  & $15 \pm 5$   & $11.1 \pm 3.7$ \\
\cline{3-6}
$ b > 40^{\circ}$ & 15 GHz & 100 $\mu$m & 0.1782  & $-11 \pm 21$ & $-1.9 \pm 3.7$ \\
\cline{3-6}
                  &        & WHAM       & 0.1354  & $51 \pm 25$  & $6.9 \pm 3.4$ \\
\cline{2-6}
                  &        & 408 MHz    & 0.6630  & $4 \pm 15$   & $2.6 \pm 9.9$  \\
\cline{3-6}
                  & 33 GHz & 100 $\mu$m & 0.2138  & $66 \pm 55$  & $14.8 \pm 11.8$  \\
\cline{3-6}
                  &        & WHAM       & 0.1499  & $47 \pm 69$  & $7.0 \pm 10.3$ \\   
\hline
\end{tabular}
}
\label{forecorrresults}  
\end{table} 
We have pictorially summarised the cross-correlation results in figure \ref{spectforeground}.
In the top row of the figure we plot the r.m.s. contribution from synchrotron to
the Tenerife data at the three Galactic cuts analysed. In black, we overplot the expected synchrotron
contribution for a spectral index of $-3.0$ and derived from the r.m.s. level in the 408 MHz
map. The contributions are systematically lower than expected at 10~GHz and the spectral index is much flatter such that for example the contribution at 15~GHz
is consistent with that at 10~GHz for $b>40^{\circ}$. We have derived
from the Tenerife data spectral indexes of $-1.77^{+1.0}_{-2.0}$, $-0.40^{+1.0}_{-1.9}$ and 
$-0.40^{+2.0}_{-1.6}$ at $b>20^{\circ}$, 
$b>30^{\circ}$ and $b>40^{\circ}$ respectively, and the r.m.s contributions based on these spectral indices is overplotted in red.

In the middle row we plot the r.m.s. contribution from free-free
emission to the Tenerife data. In black, we overplot the expected free-free emission at microwave
frequencies derived from equation \ref{halphatotempratio} which are actually in very good
agreement with the observations. In red, we plot the free-free contribution for spectral
indexes of $-0.5$, $-2.20$ and $-1.0$ at $b>20^{\circ}$, $b>30^{\circ}$ and $b>40^{\circ}$ 
respectively.

In the bottom row of figure \ref{spectforeground} we plot the r.m.s. contribution from dust 
emission to the Tenerife data which is significantly larger (few orders of magnitude) than expected from vibrational dust. In solid black we overplot
a r.m.s brightness temperature expected from the CNM model of rotational dust proposed by \cite{1998ApJ...494L..19D}. The models were rescaled to fit the data.
The intensity spectrum for this model peaks at 50~GHz which corresponds to a peak in brightness temperature around 20~GHz of
about 300 $\mu$K. From the best fit model to the data we have estimated fluctuations in the
temperature of the rotational dust of $\sim 8\%$ at the angular scales of the Tenerife
experiment. 

The r.m.s. contributions from dust to the 15~GHz data at  $b>30^{\circ}$ and $b>40^{\circ}$, as well
as the contribution from free-free at $b>20^{\circ}$ seem to be underestimated. This could be
caused by correlations between the synchrotron, dust and free-free templates which
would reduce the validity of the minimum variance solution and could bias the estimates of the
correlation.
In figure \ref{templatecorr}, we plot the following correlations, from top to bottom 
synchrotron vs dust; free-free vs dust
and free-free vs synchrotron for $b>20^{\circ}$, $b>30^{\circ}$ and $b>40^{\circ}$. We observe no correlation
between free-free and synchrotron, moderate correlation between free-free and dust
and a quite strong correlation between synchrotron and dust at low Galactic latitudes
weakening down at high latitudes. Note that at $b > 20^{\circ}$, the correlation plots
show negative vs. negative points which do not follow the correlation pattern of the
main body of points. The former points correspond to the lowest Galactic latitude data which
run into a negative beam of the triple beam pattern for the Galactic plane crossing. \\
The observed correlations between templates can not justify the observed lack of 
dust correlation at 15~GHz data at  $b>30^{\circ}$ and $b>40^{\circ}$. 
In the following we concentrate on the $b>20^{\circ}$ region for which we detect
significant dust correlated signal at all Tenerife frequencies.
 
\begin{figure*}
\begin{center}
 \includegraphics[angle=90,width=16cm,height=20cm]{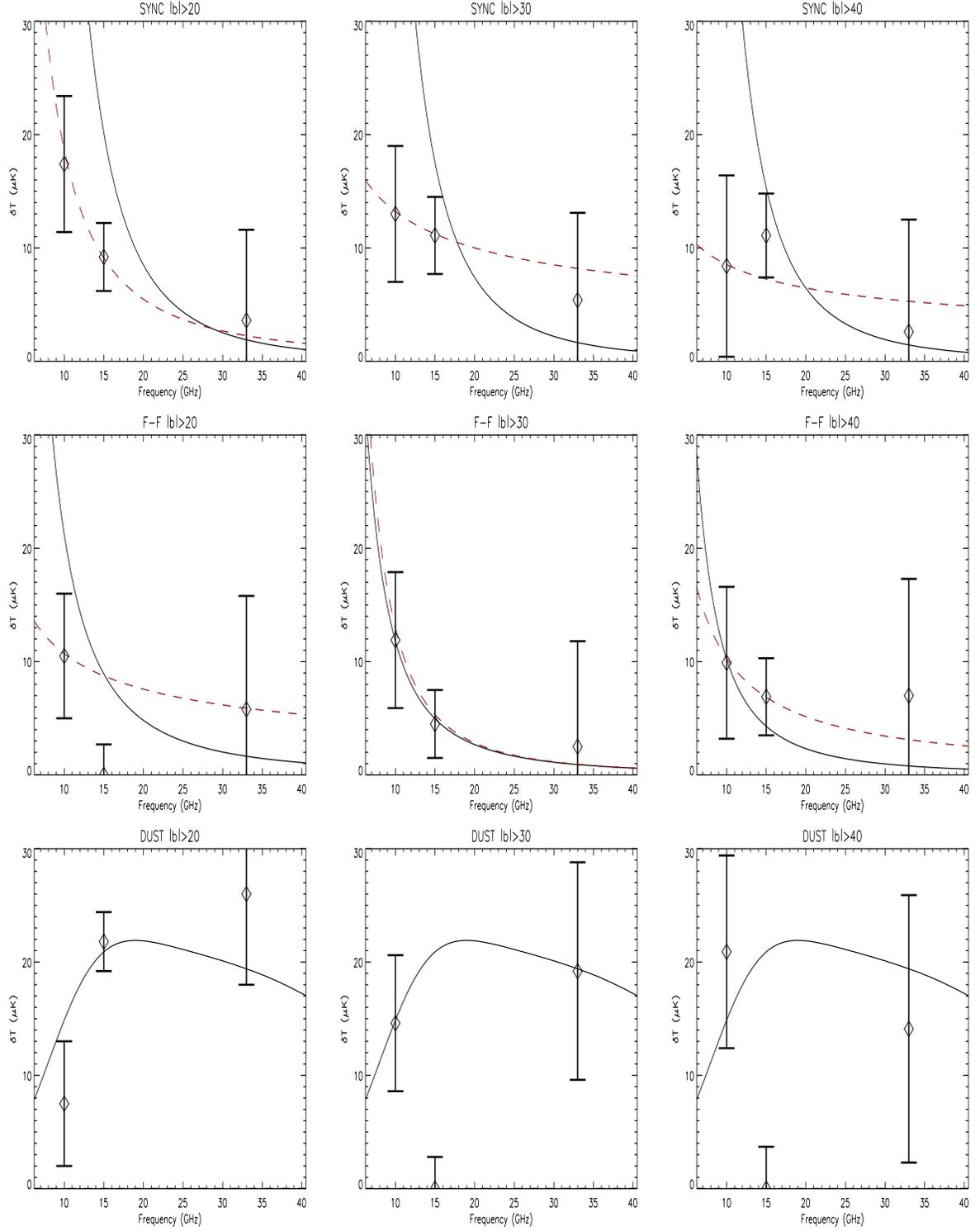}
\end{center}
\caption[{\bf Spectra of Foregrounds at the Tenerife frequencies.}]
{{\bf Spectra of Foregrounds at the Tenerife frequencies.} We represent the cross correlation coefficients as a function of frequency for
$b > 20$, $b >30$ and $b >40 $ (left, middle and right column respectively). From top to bottom, we we plot the r.m.s. contribution from synchrotron,
free-free and dust  to the Tenerife data. In solid black and red we show standard and best-fit models of the electromagnetic spectrum for each Galactic
emission. The models for dust are rescaled to fit the data. Details on the best-fit models can be found on the text.}
\label{spectforeground}
\end{figure*}

\begin{figure*}
\begin{center}
 \includegraphics[angle=90,width=16cm,height=20cm]{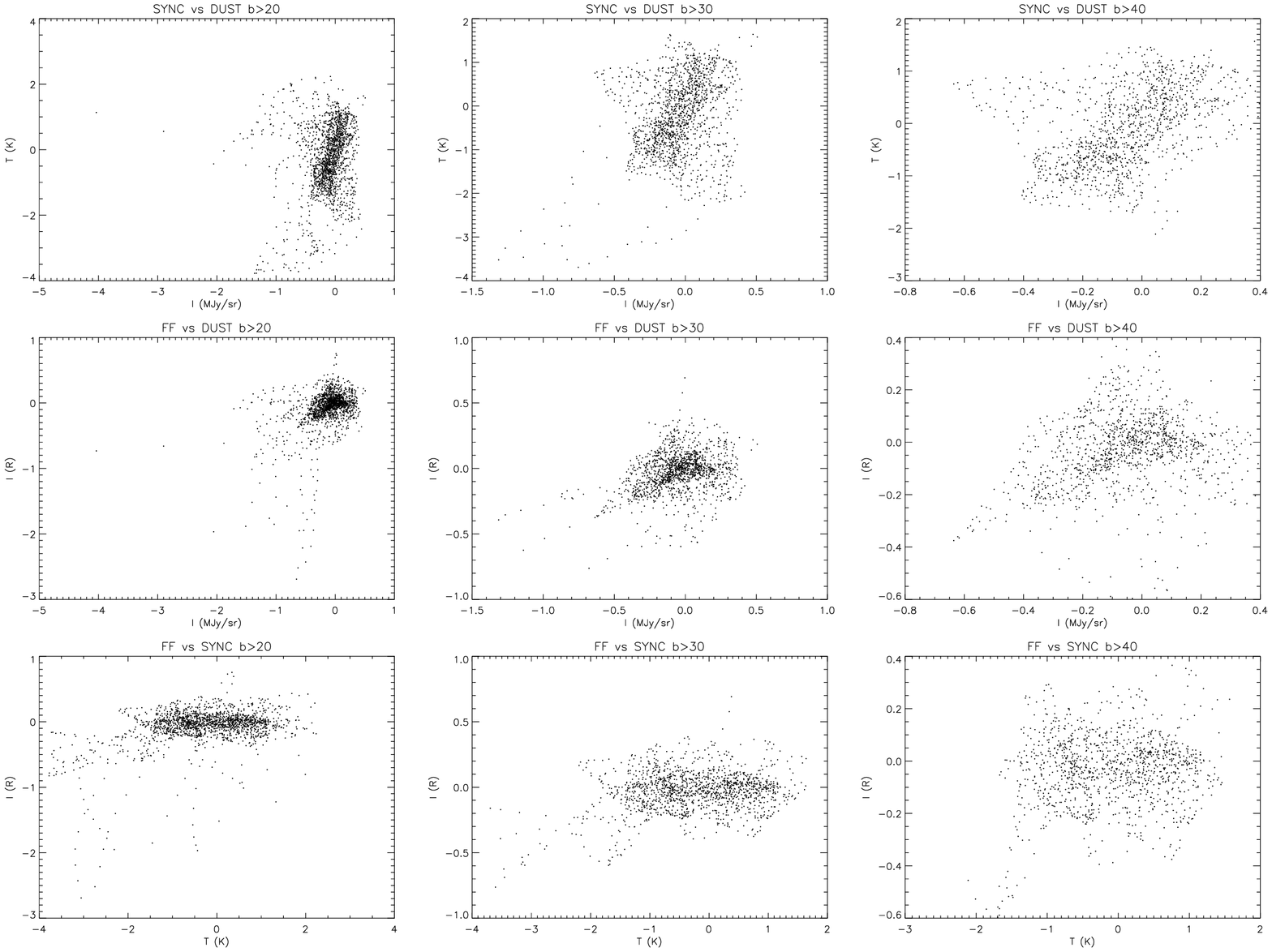}
\end{center}
\caption[{\bf Correlation between the foreground templates.}]
{{\bf Correlation between the foreground templates.}  From top to bottom, we plot synchrotron versus dust, free-free versus dust and
free-free versus synchrotron. The correlation is presented for $b > 20$, $b >30$ and $b >40 $ (left, middle and right column respectively).}
\label{templatecorr}
\end{figure*}

\section{Dust-correlated emission.}
\label{discussion}
\index{d@{\bf D}!dust correlated emission}
\cite{1996ApJ...464L...5K} cross-correlated the COBE Differential Microwave Radiometer (DMR) maps with
DIRBE far-infrared maps and discovered that statistically significant correlation did
exist at each DMR frequency, which was inconsistent with vibrational dust alone. This extra
correlation was explained at the time as free-free emission. Following this publication, 
other authors have cross-correlated CMB data sets with dust templates and found an excess of correlation which
has been interpreted as free-free emission, flat-spectrum synchrotron or emission from spinning dust. 
In table
\ref{otherexptcorr} we present an up-to-date list of CMB data for which the correlation with
dust has been performed. In all cases, a multi-template method has been used to perform the
correlation but no free-free template has been used. We have also added in this table
previous analyses of the Tenerife data by \cite{1999ApJ...527L...9D} and \cite{2001MNRAS.320..224M}
which did not include the 33~GHz data and covered a much smaller area of the sky at 10
and 15~GHz. 
\begin{table*}
\caption[Cross-Correlation results of CMB data with dust]
{{\bf Cross-Correlation results of CMB data with dust.} Available cross-correlation
coefficients of CMB data sets with dust templates at 100 $\mu$m.}
\begin{tabular}{|c|c|c|c|c|}
\hline
Experiment & Frequency (GHz) & $|b|> (degrees)$ & $\hat{a} \pm \sigma_{\hat{a}}$ $\mu$K/(MJy.sr$^{-1}$) & References \\
\hline
COBE DMR 	    &  31.5 &  20  & $18.0 \pm 2.5$    & \cite{1996ApJ...464L...5K} \\
COBE DMR	    &  31.5 &  30  & $14.5 \pm 6.0$    & \cite{1996ApJ...464L...5K} \\
COBE DMR 	    &  53.0 &  20  & $6.8 \pm 1.4$     & \cite{1996ApJ...464L...5K} \\
COBE DMR 	    &  53.0 &  30  & $6.4 3.4 \pm $    & \cite{1996ApJ...464L...5K} \\
COBE DMR 	    &  90.0 &  20  & $2.7 \pm 1.6$     & \cite{1996ApJ...464L...5K} \\
COBE DMR 	    &  90.0 &  30  & $4.6 \pm 3.9$     & \cite{1996ApJ...464L...5K} \\
\hline
Saskatoon (Ka band) &  30.0 & NCP & $15.0 \pm 8.1$    & \cite{1997ApJ...482L..17D} \\
Saskatoon (Q band)  &  40.0 & NCP & $11.8 \pm 10 $    & \cite{1997ApJ...482L..17D} \\
\hline
19 GHz survey	    &  19.0 &  20  & $38.5 \pm 3.5$    & \cite{1998ApJ...509L...9D} \\ 
19 GHz survey	    &  19.0 &  30  & $47.1 \pm 9.0$    & \cite{1998ApJ...509L...9D} \\ 
\hline
OVRO		    &  14.5 & NCP &  $209 $ & \cite{1997ApJ...486L..23L}\\
OVRO		    &  32.0 & NCP &  $36  $  & \cite{1997ApJ...486L..23L}\\
\hline
PYTHON V	    &  40.3 &       & $-3.0 \pm 18.0$   & \cite{1999ApJ...519L...5C} \\
\hline
Tenerife	    &  10.0 &  20  & $49.8 \pm 11.0 $  & \cite{1999ApJ...527L...9D} \\
Tenerife	    &  10.0 &  30  & $-8.3 \pm 31.0 $  & \cite{1999ApJ...527L...9D} \\
Tenerife	    &  10.0 &  40  & $84.0 \pm 54.0$   & \cite{1999ApJ...527L...9D} \\
Tenerife	    &  15.0 &  20  & $71.8 \pm 4.5$    & \cite{1999ApJ...527L...9D} \\
Tenerife	    &  15.0 &  30  & $94.9 \pm 15.0 $  & \cite{1999ApJ...527L...9D} \\
Tenerife	    &  15.0 &  40  & $72.0 \pm 33.0$   & \cite{1999ApJ...527L...9D} \\
\hline
Tenerife	    &  10.0 &  20  & $71.0 \pm 18.0$   & \cite{2001MNRAS.320..224M} \\
Tenerife	    &  10.0 &  30  & $-7.0 \pm 32.0$   & \cite{2001MNRAS.320..224M} \\
Tenerife	    &  10.0 &  40  & $28.0 \pm 39.0$   & \cite{2001MNRAS.320..224M} \\
Tenerife	    &  15.0 &  20  & $91.0 \pm 11.0$   & \cite{2001MNRAS.320..224M} \\
Tenerife	    &  15.0 &  30  & $29.0 \pm 20.0$   & \cite{2001MNRAS.320..224M} \\
Tenerife	    &  15.0 &  40  & $3.0 \pm 26.0$    & \cite{2001MNRAS.320..224M} \\
\hline
South Pole 94	    &  30.0 & 40      & $20.0 \pm 36.0$   & \cite{2001AA...368..760H} \\
South Pole 94	    &  40.0 & 40      & $68.1 \pm 42.4$   & \cite{2001AA...368..760H} \\
\hline
\end{tabular}
\label{otherexptcorr}
\end{table*}

The correlation coefficients we deduced are significantly smaller at $|b| > 20$
than those calculated by \cite{1999ApJ...527L...9D} and \cite{2001MNRAS.320..224M}. This is probably
due to 50 \% larger area (localized signals are diluted) and a more careful subtraction of baselines at low Galactic latitudes 
in our data.  The inclusion of an extra free-free template seems to play minor role on this at 10~GHz and none at 15~GHz.\\

The cross-correlation results for the Tenerife experiment presented in this paper
confirm the existence of extra dust-correlated emission at microwave frequencies and suggest that
it is not due to free-free emission. 
The moderate correlation found between the dust and
free-free templates could confuse the results obtained but clearly cannot account
for all the observed dust-correlated component. A more detailed study is needed to take
into account correlations between templates in the calculation of the correlation coefficients.

In figure \ref{b20corrdata} we plot the dust cross-correlation coefficients ( $|b| > 20$) for 
the Tenerife (diamonds) and COBE-DMR (triangles) data. The rest of the experiments presented 
in table \ref{otherexptcorr} were not included in this plot because they
observe at different angular scales and different areas of the sky (see table for details)
and therefore no direct comparison with the Tenerife and COBE 
data is possible. The Tenerife data suggest a peak in the spectrum at about 30 GHz although
this is mainly based on the data at 33~GHz which is significantly much noisier
and cover a much smaller part of the sky.
However DMR data combined with the 10 and 15 GHz Tenerife data points prefer a peak
in the range 15-20~GHz. 
In color, we overplot the six
spinning dust models proposed by \cite{1998ApJ...494L..19D} combined with the standard vibrational
dust model described in the previous sections, for which the amplitude has been taken as a free 
parameter and fitted to the COBE and Tenerife 10 and 15~GHz data. We observe that the shape of
the spectrum given by the data is similar to some of the predicted spectra although the models 
do not indicate a sharp rise in the range 10-15~GHz. 

\begin{figure}
\begin{center}
  \includegraphics[width=9cm]{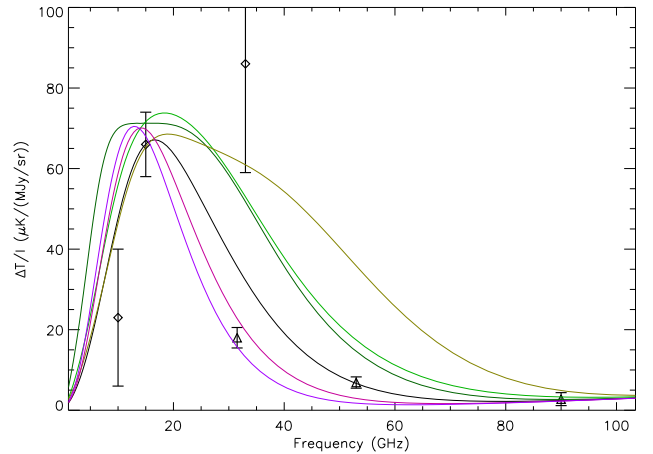}
\end{center}
\caption[{\bf Cross-correlation coefficients for the dust-correlated emission.}]
{{\bf Cross-correlation coefficients for the dust-correlated emission.}
We plot the Tenerife data as diamonds and the COBE DMR data as triangles. The
color solid curves overplotted correspond to joint vibrational and rotational dust
models as described in the text.}
\label{b20corrdata}
\end{figure}

\section{Conclusions}
\label{conclusion}
We have presented in this paper a re-analysis of the Tenerife data including previously
published data and new data from January 1998 to December 2000. 
This analysis leads to evidences for an excess of dust-correlated
emission at microwave frequencies in the range 10--33~GHz and at
large angular scales, from 5 to 15 degrees. This correlation can not all be associated with 
free-free emission. A combination of the Tenerife and COBE data suggests spinning dust emission
could account for the extra correlation. However, the scatter observed in the
data and the discrepancy in the spectrum shape indicate other components may also be
responsible for the extra correlation. 
Furthermore, the analysis does not take into account correlations between the Galactic emission templates and this can bias the estimate of the correlation coefficients.
To correct for this a more detailed analysis, which will account for the cross-correlation between templates, is needed.

\section*{acknowledgements}
The authors would like to thanks the editors of this special issue.

\bibliographystyle{juan} 
\bibliography{mybib,cmbpapers,radiogeneral,biblio} 

\end{document}